# The Paradox of the Recoil Force Acting on a Leaking Water Tank

Avi Marchewka and Yiftah Navot

Eldad Six-Year School 901 Golda Meir Boulevard Netanya 4216101Israel

Abstract

In the first part of the article, we will outline the paradoxical picture that arises when attempting to calculate the recoil force of the water leakage from a hole at the bottom of a water tank. We will present three different options for the recoil force acting on the water tank as a result of this leakage (in Chapters 1, 2 and 3). In the second part of the article, we will present an experiment that resolves this question (in Chapter 4). Finally (in Chapter 5), we will present a coherent picture of the description of the leakage and the result recoil force.

## Introduction

The main question addressed in this article is the existence of recoil forces due to the vertical flow of water through a hole at the bottom of a water tank, as depicted in Figure 1.

The recoil force is the driving force in rocket propulsion. The recoil force is also evident in shooting a rifle and in physics problems about leaking water carts. In these systems, the recoil force is sometimes accompanied by an internal force in the system, such as fuel combustion in a rocket or the explosion of gunpowder in a bullet. The internal force expels part of the system's mass, causing a reaction force according to Newton's third law. The reaction force is the recoil force.

In the case of vertical flow, it is not immediately clear whether such an internal force exists, and if it does, what its source is. In Chapter 1, we present the question and three possible answers. In Chapter 2, we calculate the velocity of the water flow and the rate of decrease of the water level in the tank. In Chapter 3, we provide plausible but not definitive explanations for each of the possible answers. Since the theoretical models and calculations do not fully overlap and give different results, we do not have a definitive answer to the question of whether a recoil force exists in this case. So, in Chapter 4, we turn to experiment. In Chapter 5, we explain the discrepancies between the different calculations and deepen our understanding of the recoil force.

## Chapter 1: Presentation of the Question

Consider a water tank with a leaking from a hole at the bottom, as depicted in Figure 1.

**Question:** What is the recoil force that the water flow exerts on the water tank?

**Tentative Answers to be considered:**

A.  $F_{recoil} = 0$

B.  $F_{recoil} = \rho a g h$

C.  $F_{recoil} = 2\rho a g h$

where $\rho$ is the density of the water, $a$ is the area of the hole, $h$ is the height of the water level in the tank, and $g$ is the acceleration due to gravity.

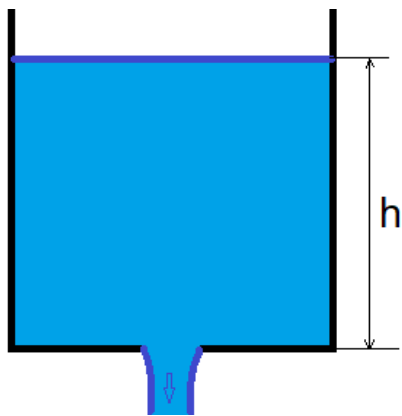

Figure 1: Water leaking from the tank

Next, we derive the basic laws that we make use of: Torricelli's law and the time dependency of the water level height.

## Chapter 2: The flow out of the leaking tank

**2.A:** Torricelli's law

Let's assume that the area of the hole is much smaller than the cross-sectional area of the tank. According to Bernoulli's principle,

$$P_0 + \frac{1}{2}\rho u^2 = P_0 + \rho g h \quad (1)$$

where $P_0$ is the atmospheric pressure and u is the velocity of the water flow exiting the tank.

From this equation, we derive Torricelli's law:

$$(2) \qquad u = \sqrt{2gh}$$

Note that the velocity $u$ is equal to the velocity that a body would reach in free fall under acceleration g from a height $h$. However, we want to emphasize that this velocity is generated by the pressure at the bottom of the water tank, and not directly by gravity.

**Further note:** In this calculation, we assume that the hole is small relative to the cross-sectional area of the tank, and thus we neglect the velocity of the water flow within the tank compared to the exit velocity. A more detailed derivation is presented in Appendix A.

## 2.B: Time dependence of the height of the water

From the principle of conservation of mass flux, there is a relationship between the velocity of the water exiting the tank ($u$) and the rate of decrease of the water level in the tank ($v$):

$$(3) \qquad Av = au$$

where $A$ represents the cross-sectional area of the tank and $a$ represents the cross-sectional area of the hole.

Using Torricelli's law (Eq. (2)), the equation for the rate of decrease of the water level in the tank is:

$$(4) \qquad v = \frac{a}{A} u = \frac{a}{A}\sqrt{2gh}$$

From this, we obtain a differential equation for the height of the water level in the tank as a function of time,

$$(5) \qquad \dot{h} = -\frac{a}{A}\sqrt{2gh}$$

where the dot above the $h$ denotes differentiation with respect to time, and we used the fact that $\dot{h} = -v$.

The solution to this equation with the initial condition $h(0) = h_0$ is parabolic in time:

$$(6) \qquad h(t) = \left(\sqrt{h_0} - \frac{1}{2}\frac{a}{A}\sqrt{2g}\, t\right)^2$$

The minimum point of the parabola is reached when the height is zero, i.e., when the tank is empty.

In the experiment described later, we also examined the fit of the water level height graph to this formula.

**Note:** Due to what is known as the vena contracta effect [2], the effective hole area that is slightly smaller than the physical hole area. Richard Feynman's heuristic explanation [1] for this correction is that the flow exiting the hole has a horizontal component, which cancels out slightly below the height of the hole, and the cross-sectional area of the stream at this height is the determining factor.

# Chapter 3: Justifying arguments for each of the three answers for the question in chapter 1

Next, we argue why each of the above answers might be correct, depending on what exactly is included in the theoretical model.

### 3.A Why Answer A could be correct

Let's look at a simple example: A person holding a ball stands in a room on top of a tower, with a hole in the floor of the room. The person lifts their hand two meters above the floor and releases the ball. The released ball falls down and exits through the hole in the floor.

**Question:** Did the ball exert a recoil force on the person or the room at the top of the tower?

From an external observer's perspective, it seems that the ball exited the floor at a speed of more than 6 meters per second, so they might think that the ball was thrown down with force and exerted a recoil force on the body that threw it. However, the person simply released the ball, not throwing it down with force, so no recoil force acted on them.

Similarly, let's assume that the tank with the hole has an internal structure that includes a pipe extending from the hole at the bottom to close to the water surface, as depicted in Figure 2.

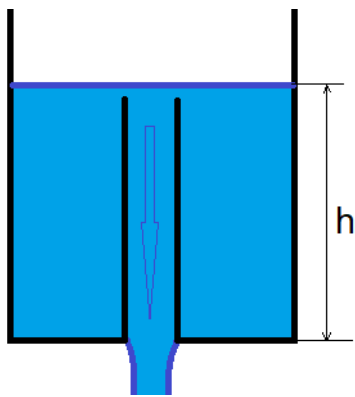

Figure 2: The internal structure of the tank with the internal pipe

For an external observer, at least momentarily, there is no difference between the system in Figure 1 and the system in Figure 2. According to Torricelli's law (Equation (2)), the exit velocity of the water is the same in both cases. However, like in the case of the person who releases a ball at the top of the tower, the water exiting the hole in the system of Figure 2 does not create any recoil force. Therefore, Answer A ($F_{recoil}= 0$) is correct.

### 3.B.1 Why Answer B could be correct

To understand the recoil force, let's consider how we might measure it. One way to do this is to suspend the tank with water from a scale (or a force gauge), as depicted in Figure 3.

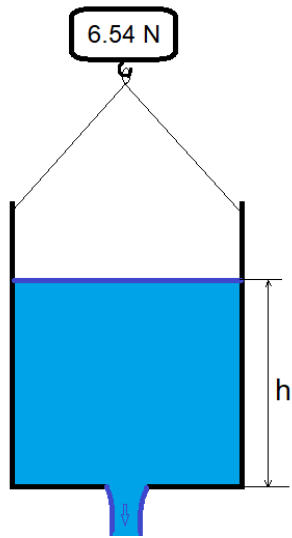

Figure 3: Water leaking from a tank suspended on a scale

Now we can obtain the water level in the tank and the weight of the tank with the water at any moment during the experiment. According to the water level, we can calculate the mass of the water in the tank based on the water density multiplied by the volume,

$$(7) \qquad M_w = \rho A h$$

Assuming a quasi-steady state (small hole), we can calculate the recoil force based on the balance of forces,

$$(8) \qquad F_{recoil} = (M_c + M_w)g - W_M$$

where $M_c$ is the mass of the empty tank, and $W_M$ is the reading shown by the force gauge.

Next, assuming that the tank with the hole (Figure 1) is equivalent to the tank with the internal pipe (Figure 2), we replace the tank in Figure 3 with the tank with the internal pipe from Figure 2. Now, the water currently inside the internal pipe is in free fall and thus not weighed by the force gauge.

Therefore,

$$(9) \quad W_M = (M_c + M_w)g - \rho ahg$$

And thus,

$$(10) \quad F_{recoil} = \rho ahg$$

**Conclusion:** Answer B ($F_{reciol} = \rho ahg$) is correct.

**Note:** It is not entirely clear whether it is appropriate to call the weight difference obtained here a recoil force. However, if we return to the seemingly equivalent system of the simple tank (without the internal pipe), it will at least appear as a recoil force.

### 3.B.2 A second argument for Answer B

Let us reconsider the system described in Figure 1 (and in Figure 3). It seems that the water inside the tank flows very slowly toward the hole. However, near the hole, the ground falls away, and the water pressure inside the tank exerts a force that accelerates the water to the exit velocity.

According to Bernoulli's equation, the pressure difference between the inside and the outside of the hole is

$$\Delta P = \rho g h$$

The force that accelerates the water is equal to the pressure difference multiplied by the area of the hole, $a$:

$$(11) \quad F = \rho g h a$$

According to Newton's third law, the force that creates the downward water flow comes with an equal size and opposite direction force acting on the tank with the water: this is the recoil force.

Therefore, once again, we find that Answer B ($F_{recoil} = \rho g h a$) is correct.

### 3.B.3 A third argument for Answer B: Calculation of the Recoil Force Using Work and Energy Considerations

Let's revisit the flow throughout the hole from energy perspective. As described in the previous section, near the hole the ground falls away, and the water pressure inside the tank exerts a force that accelerates the water to the exit velocity. However, it is not clear that the effective area on which the pressure acts is the hole area, as the water flows to the hole from the sides as well. Therefore, we take a different approach.

In order to calculate the recoil force, we will compute the energy gained by the water entering the jet exiting the hole over a small time interval $\Delta t$. The mass of the water exiting the tank during this time is given by the water density multiplied by the volume of the exiting water,

$$(12) \qquad \Delta m = \rho \cdot a \cdot u \Delta t$$

This water element starts with a velocity close to zero and, under the influence of the water pressure at the bottom of the tank, reaches velocity $u$. Therefore, the change in its kinetic energy is given by

$$(13) \qquad \Delta E = \frac{1}{2} \Delta m \cdot u^2 = \frac{1}{2} \rho \cdot a \cdot u^3 \cdot \Delta t$$

This change in the kinetic energy should be equal to the work of the driving force along an acceleration zone of length $\Delta x$,

$$(14) \qquad \Delta E = F \cdot \Delta x = F \cdot u \Delta t$$

Notice that for the length of the acceleration path inside the thank, $\Delta x$, we take the length of the water travel outside the tank, $u\Delta t$. We will return to this assumption below.

By substituting $\Delta E$ from Eq. (13) and divide by $u\Delta t$ we get

$$(15) \qquad F = \frac{1}{2} \rho \cdot a \cdot u^2$$

Using Torricelli's law (Equation (2)), we find:

$$(16) \qquad F = \frac{1}{2} \rho \cdot a \cdot 2gh = \rho g h a$$

According to Newton's third law, the recoil force is equal to the force that accelerated the water.

Therefore, Answer B is correct (for the third time).

### 3.C Why Answer C is Correct

As described in section 3.B.2, near the hole, the ground falls away, and the water pressure inside the tank exerts a force that accelerates the water to the exit velocity. Again, it is not clear what is the effective area on which the pressure acts, and again, we will take a yet different approach.

To calculate the recoil force, we will compute the momentum gained by the water entering the jet exiting the hole over a short time interval $\Delta t$. The mass of the water exiting the tank during this short time is given by the water density multiplied by the volume of the exiting water as described by equation (12) above.

The water element starts with a velocity close to zero and, under the influence of the water pressure at the bottom of the tank, reaches velocity $u$. Therefore, its change in momentum is

$$(17) \qquad \Delta p = \Delta m \cdot u = \rho \cdot a \cdot u^2 \cdot \Delta t$$

Consequently, the force that creates the water jet, as obtained from the derivative of the momentum with respect to time, is

$$(18) \qquad F = \frac{\Delta p}{\Delta t} = \rho \cdot a \cdot u^2$$

Using Torricelli's law (Equation (2)), we find,

$$(19) \qquad F = \rho \cdot a \cdot 2gh$$

According to Newton's third law, the recoil force is equal to the force that accelerated the water.

Therefore, Answer C ($F_{recoil} = 2\rho gha$) is correct.

## Chapter 4: The Experiment and Results

The models and their analysis in the previous chapters did not lead to a definitive conclusion. The paradoxical situation of theory means that it makes sense to turn to experiment. The experiment described in Chapter 3.B.1 was carried out by the students of Avi Marcherwka, Arie Upar and Nave Levi Zadok, using a tank made from a transparent cylindrical tube with an internal diameter of 18 mm and a length of 1.5 meters, with holes of different diameters: 4 mm, 7 mm, and 10 mm. The weight during the experiment was measured using a PASCO force gauge, and the height was measured by filming the experiment and tracking the water level in the video using the Tracker application.

**Graph 1** shows the water level as a function of time compared to the theoretical formula (Equation (6)), for a hole with a diameter of 4 mm.

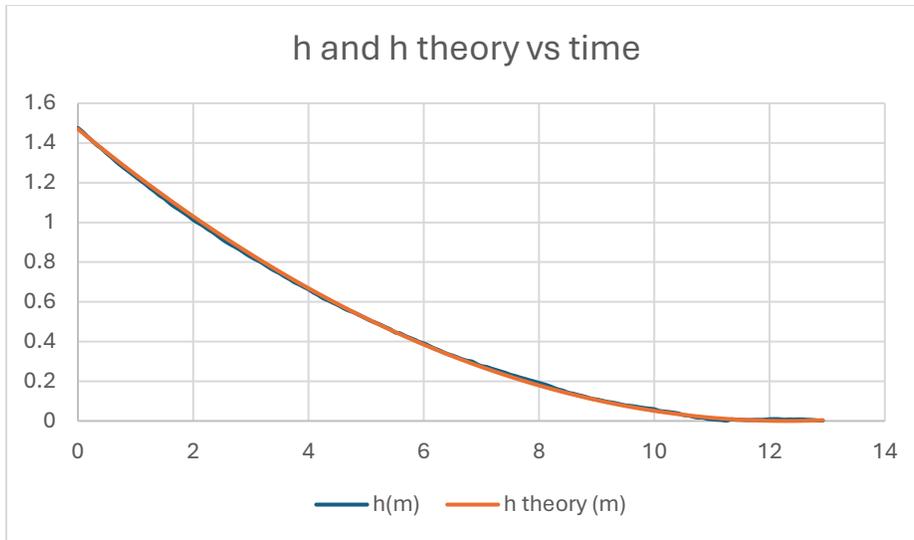

Graph 1: the water level vs. time for a 4 mm hole. In blue is the measurement, and in orange is the theoretical calculation.

Note: According to the note in section 2.B, the effective area of the orifice is slightly smaller than the physical area of the orifice. Accordingly, for optimal alignment, the calculation of the water level height according to equation (6) is done based on an orifice with an effective diameter of 3.8 mm instead of 4 mm.

The following graph presents the results of measuring the weight compared to calculating the weight according to the water surface height for the 4 mm diameter orifice.

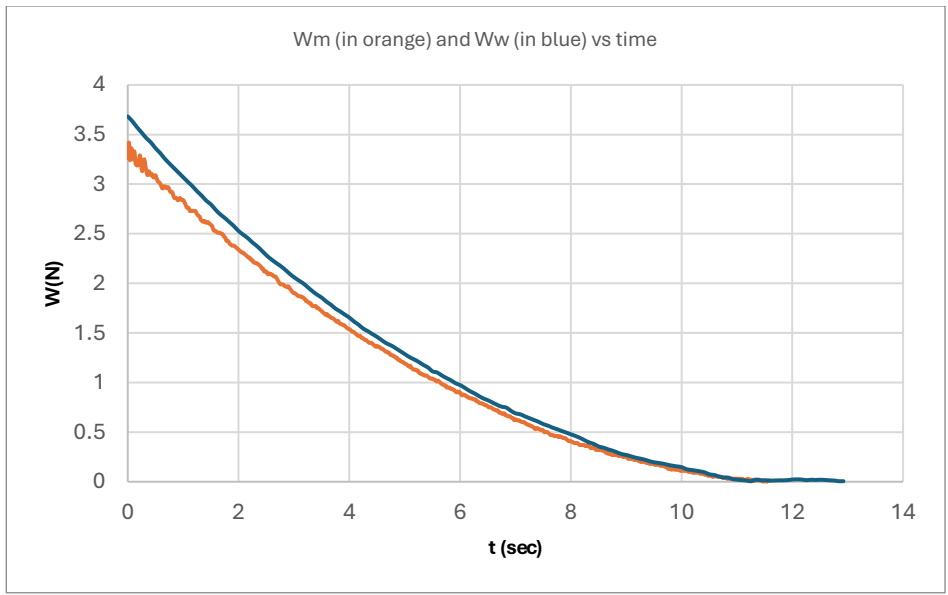

Graph 2: The calculated water weight according to the height measurement (in blue) and the weight according to the force gauge (in orange) as a function of time.

Graph 2 illustrates the discrepancy between the two lines. The gap between them represents the recoil force. This gap indicates that answer A in the question posed in Chapter 1 is not correct.

Graph 3 presents the comparison to the theoretical graph, where the recoil is calculated according to the formula provided in answer C to the question presented in Chapter 1

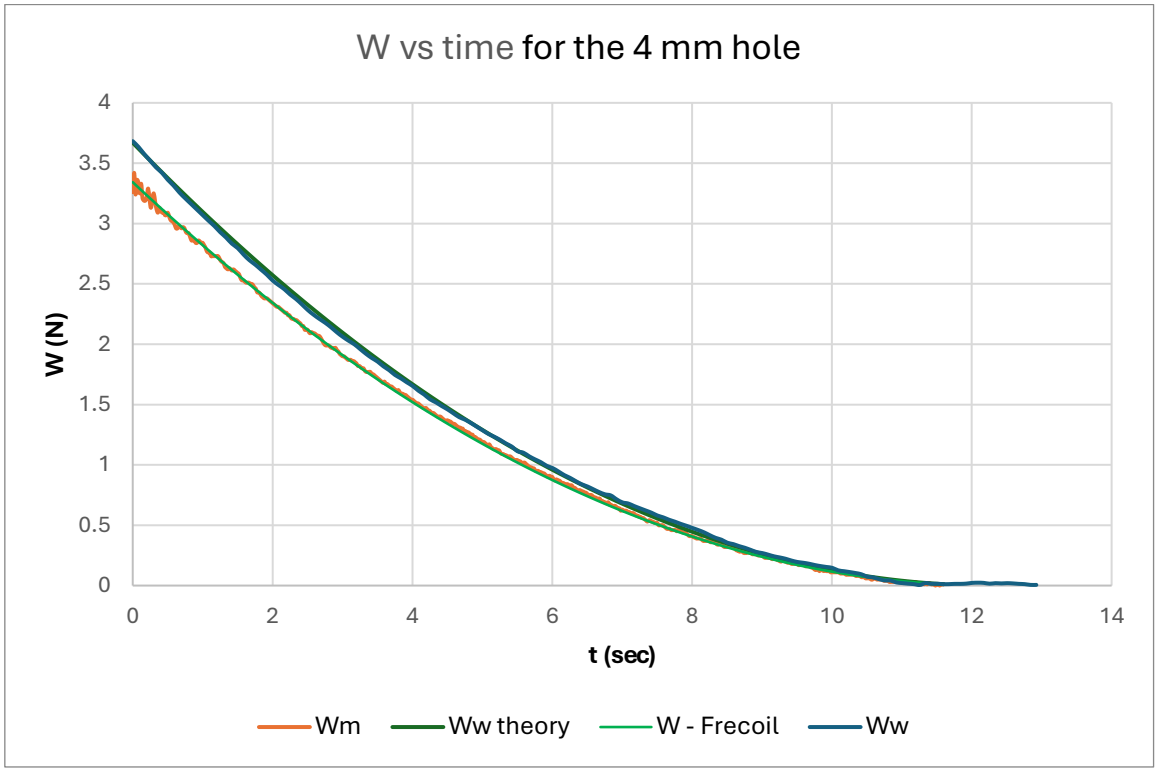

Graph 3: Comparison to theory for a 4 mm diameter orifice. In orange: the weight measurement, in blue: the weight calculated based on the height measurement, in green: the weight calculated minus the recoil force, in gray: the weight calculated based on the calculated height.

As seen in the graph, there is a good match between the measurement of the force meter (in orange) and the calculation (in green). The unavoidable conclusion is that there is recoil force, as per Answer C.

$$(20) \quad F_{recoil} = 2\rho a g h$$

Note: Here as well, the calculation of the recoil force is based on a hole with an effective diameter of 3.8 mm.

The results of the experiments with holes of diameters 7 mm (with an effective diameter of 6 mm) and 10 mm (with an effective diameter of 8.8 mm) yielded similar matches, as shown in the following graph 4 and graph 5.

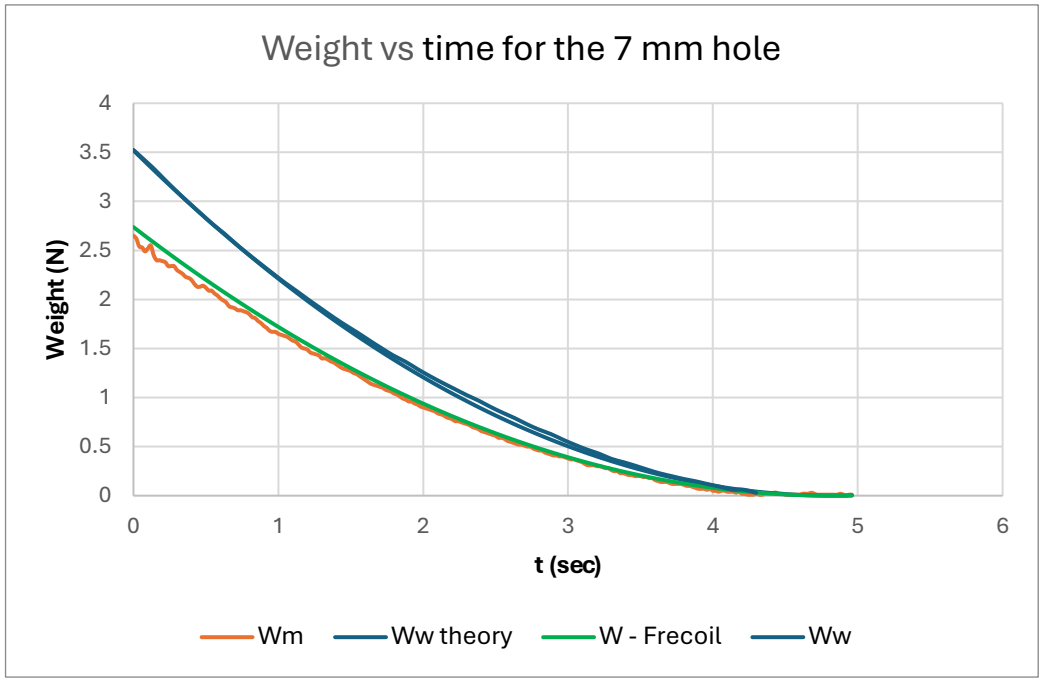

Graph 4: Comparison to theory for a 7 mm diameter orifice. In orange: the weight measurement, in blue: the weight calculated based on the height measurement, in green: the weight calculated minus the recoil force, in gray: the weight calculated based on the calculated height.

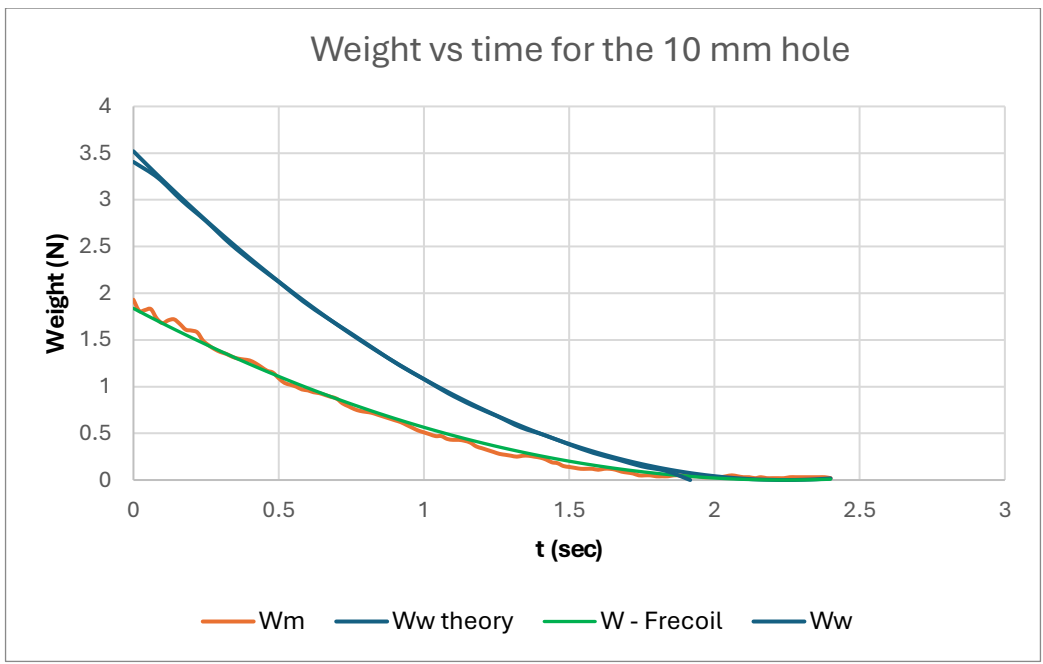

Graph 5: Comparison to theory for a 10 mm diameter orifice. In orange: the weight measurement, in blue: the weight calculated based on the height measurement, in green: the weight calculated minus the recoil force, in gray: the weight calculated based on the calculated height.

# Chapter 5: Discussion of the Contradictory Physical Models Presented in Chapter 3

The experiment showed a recoil force, in agreement with Answer C to the question in Chapter 1, and by Eq. (20),

$$F_{recoil} = 2\rho agh$$

In Chapter 3, several different physical arguments lead to different results for the recoil force. We can see that the arguments based on the behavior of the water flow inside the tank give a wrong answer. It seems that the detailed water flow inside the tank should not be neglected. We will attempt to understand the errors in the explanations that led to the incorrect results.

## 5.A: Errors in the arguments 3.A and 3.B.1

Let's examine the flow of water within the internal pipe in Figure 2. Since the water flow starts at a very low speed at the top and increases its speed due to gravity when descending inside the pipe, and since the water flux at every height should be constant, it appears that the diameter of the water flow within the internal pipe should become thinner as it reaches the exit at the hole, as described in Figure 4. Therefore, the calculations presented in sections 3.A and 3.B.1 are based on a flawed analogy.

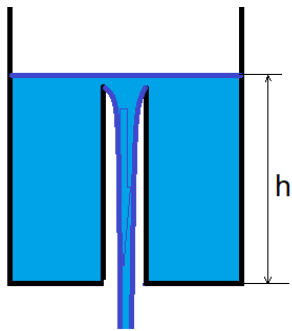

Figure 4: depicting water dripping from a container with an internal pipe

Note that this argument might be corrected by replacing the cylindrical internal pipe with a funnel shaped pipe that follows the shape of the thinning flow. As argued in the original argument in Section 3.B.1, we must subtract the weight of the free-falling water from the total weight. Only the volume of the free-falling water is now the volume of the funnel instead of the volume of the cylindrical pipe. As would be shown below (see Eq. (24)), the average cross sectional area of the funnel is twice the area of the hole, and therefore the volume of the funnel is twice the volume of the cylindrical pipe. Therefore, the resulting total weight of the free-falling water with the internal funnel shape pipe converges to answer C.

## 5.B: What's wrong in the Work and Energy argument presented in Section 3.B.3

It appears that the incorrect assumption in the calculation is hidden in equation (14). In this equation we take $\Delta x = u\Delta t$ for the calculation of the work. However, if the small water element starts from a velocity close to zero, and ends with velocity $u$, then its average speed is only $v_{avg} = \frac{1}{2}u$, and the distance (on which the force is acting along) becomes:

$$(21) \qquad \Delta x = \frac{1}{2}u\Delta t$$

Therefore, the right equation for the driving force (instead of equation (14)) is

$$(22) \qquad \Delta E = F \cdot \Delta x = F \cdot \frac{1}{2}u\Delta t$$

And by substituting $\Delta E$ from Eq. (13), we get the correct answer (Answer C):

$$F_{recoil} = 2\rho gah$$

However, if we go back to the beginning of the argument of this Section 3.B.3, we find another use of $\Delta x = u\Delta t$ in equation (12), when we take $\Delta m = \rho \cdot a \cdot \Delta x = \rho \cdot a \cdot u\Delta t$.
The following question comes up: Should we change the calculation of $\Delta m$ as well?
The (slightly surprising) answer is No.

To understand this, we go back to the law of flux conservation along the acceleration zone. According to this law, at each point along the flow line, the area of the flow cross-section times the velocity of the flow at that point should be constant:

$$S(z) \cdot v(z) = Const$$

were $z$ is a parameter along the flow line and $S(z)$ is the cross-section area and $v(z)$ is the velocity of the flow at the point z. At the end of the acceleration, we have $S(z_e) = a$ and $v(z_e) = u$ and therefore we get:

$$(23) \qquad S(z) \cdot v(z) = au \quad \forall z$$

Accordingly, as the velocity grows along the acceleration zone, the cross-section area should be reduced. And thus, the flow zone does not look like a cylindrical pipe. It looks more like a funnel (see Figure 5). In particular, the flux along the acceleration zone stays $a \cdot u$, and Eq. (12) remains correct: $\Delta m$ is equal to the density of the water times the flux times $\Delta t$.

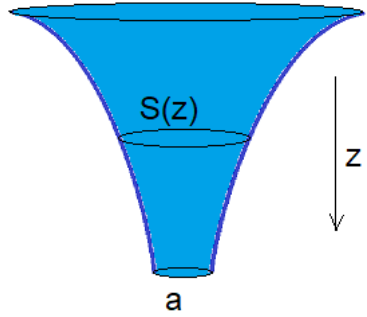

Figure 5: the flow funnel

Note also that as the average flow velocity is half the final velocity $u$, the average cross section area should be twice the area of the hole in order to keep same the total flux. Namely, as $A_{avg} \cdot v_{avg} = au$, and as $v_{avg} = \frac{1}{2}u$, we get the following:

$$(24) \qquad A_{avg} = 2a$$

As discussed above, this can resolve the mistake in argument 3.B.1. And, as will be discussed below, this also resolves the mistake in Argument 3.B.2.

### 5.C: Resolving the mistake in the calculation presented in Section 3.B.2

In Section 3.B.2, the calculation of the recoil force is based on the area of the hole multiplied by the pressure difference. But, as explained above, the effective average area of the flow along the acceleration zone is twice the area of the hole. Therefore, the total force becomes double the result presented in Eq. (11), and so the result recoil force converges to Answer C as well.

In Appendix B, we give a generalization of the concept of an effective area for the action of the pressure difference.

### 5.D: Examine the Shape of funnel

In the assumption of constant acceleration along the acceleration zone, we can use the equation of constant flux (Eq. (22)) to calculate the cross-sectional area as a function of length $z$. As the water element starts with a small velocity $v_0$ and ends with velocity $u$, the constant acceleration should be

$$(25) \qquad \tilde{g} = \frac{u - v_0}{\Delta t} \cong \frac{u}{\Delta t}$$

Were $\Delta t$ is the acceleration time.

The velocity as a function of the distance at a motion of a constant acceleration, and a small initial velocity $v_0$ is given by:

$$(26) \quad v(z) = \sqrt{v_0^2 + 2\tilde{g}z}$$

And therefore, the cross-section area is given by

$$(27) \quad S(z) = \frac{au}{\sqrt{v_0^2 + 2\tilde{g}z}}$$

The total volume in the acceleration funnel becomes

$$V_{funnel} = \int_0^{z_e} \frac{au}{\sqrt{v_0^2 + 2\tilde{g}z}} dz = \left[\frac{2au}{2b}\sqrt{v_0^2 + 2\tilde{g}z}\right]_0^{z_e} = \frac{au^2}{\tilde{g}} - \frac{auv_0}{\tilde{g}}$$

For a very small initial velocity, substituting $\tilde{g}$ from Eq. (25) and substituting $\Delta x$ from Eq. (21) we obtain,

$$V_{funnel} \cong \frac{au^2}{\tilde{g}} \cong au\Delta t = 2a\Delta x = 2 \cdot V_{cylinder}$$

Indeed, we found that the volume of the funnel is twice the volume of the cylinder of height $\Delta x$ above the area of the hole.

**Note about the spatial shape of the acceleration funnel:** In the actual flow in the water tank, the surfaces of constant flow speed (which are the surfaces of constant parameter $z$), should not be on a spatial plane, as described in Figure 5. The maybe more realistic spatial shape of the acceleration funnel is shown in Figure 6.

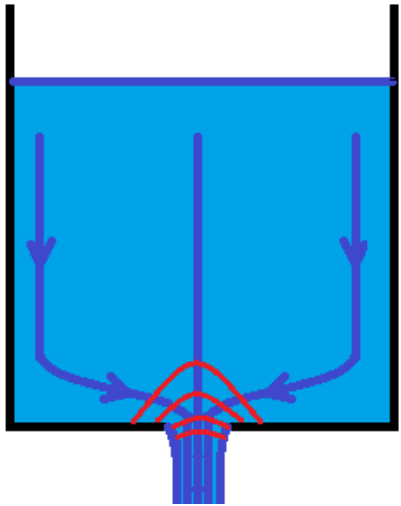

Figure 6: the stream lines in blue and the spatial shape of the surfaces of same flow speed in red

## Summary and discussion

We discussed the recoil force acting on a container of water with a hole at the bottom. Three possibilities for the magnitude of the recoil force were presented, along with five calculations based on different physical considerations. To determine what is the right answer, whether there is a recoil force and what its magnitude is, we conducted an experiment in which we simultaneously measured the height of the water level and the weight of the container with the water throughout the process of water exiting the container.

The conclusion from the experiment is that the recoil force is given by: $F_{recoil} = 2\rho a g h$.

From Chapter 3, it can be observed that the arguments that lead to Answers A and B for the question in Chapter 1 were derived by analyzing the state of the water inside the container, whereas the argument that leads to Answer C was derived from analyzing the outgoing water. It appears that the source of the error in the wrong calculations of the recoil force arises from misconception of the water flow at the acceleration zone inside the tank. Indeed, in Chapter 5 (and in Appendix B), we showed that the wrong arguments might be corrected by considering a funnel shaped flow at the acceleration zone.

Furthermore, this funnel shaped flow at the acceleration zone leads to an interesting conclusion: it appears that the internal pressure accelerating the water exiting the container acts on an effective area twice as large as the orifice area. Note that a more accurate understanding might be obtained by solving the Navier-Stokes equations, or, in the simpler case, the Euler equations for this flow problem.

The experiment described in Chapter 4 is simple to perform and requires basic equipment, with easy-to-measure results. It can be used to teach undergraduate physics students the equations of motion for dynamic systems with variable mass. This experiment is also suitable for engineering students studying fluid dynamics and Bernoulli's equation.

**Acknowledgement**

The author thanks the students Arie Upar and Nave Levi Zadok for conducting the experiments. Special thanks to the head of the Sharet High School lab, where the experiment took place, Hayim Ehlers Vilela, for his creative help in the experiments. A.M. wishes to thank Oskar Pelc for fruitful discussions.

**References**

1. *Feynman, Richard P. (Richard Phillips), 1918-1988. The Feynman Lectures on Physics. Reading, Mass. :Addison-Wesley Pub. Co., 19631965.*
2. Gerhart, Philip, M. et al. Munson, Young and Okiishi's Fundamentals of Fluid Mechanics. Available from: VitalSource Bookshelf, (8th Edition). Wiley Global Education US, 2018.

**Appendix A: Calculation of the flow velocity again**

Torricelli's law is derived under the assumption that the velocity of the water's descent within the container is negligible compared to the exit velocity. We can refine this consideration in Bernoulli's equation for the case where the orifice is still small, allowing us to assume that the flow is momentarily steady, but the velocity of the water's descent in the container is not completely negligible. According to Bernoulli's principle, in this case, we get:

$$(A1) \quad P_0 + \frac{1}{2}\rho u^2 = P_0 + \frac{1}{2}\rho v^2 + \rho g h$$

And according to the relationship between $v$ and $u$ presented Equation (3), we get:

$$(A2) \quad P_0 + \frac{1}{2}\rho u^2 = P_0 + \frac{1}{2}\alpha^2 \rho u^2 + \rho g h$$

Where $\alpha$ denotes the area ratio: $\alpha = a/A$.

From here we get:

$$(A3) \quad u = \sqrt{\frac{2gh}{1-\alpha^2}}$$

In other words, according to this calculation, there is a correction to Torricelli's law (Equation (2)) by a factor of $\frac{1}{\sqrt{1-\alpha^2}}$.

The following table presents the magnitude of this correction under the experimental conditions described in Chapter 4:

| d (mm) | d effective (mm) | D (mm) | Alpha | 1/sqrt(1-alpha^2) |
|---|---|---|---|---|
| 4 | 3.8 | 18 | 0.0446 | 1.0010 |
| 7 | 6 | 18 | 0.1111 | 1.0062 |
| 10 | 8.8 | 18 | 0.2390 | 1.0298 |

Table A1: Correction Factor to Torricelli's Law for the Three Orifices in the Experiment
As shown in the table, even for the largest orifice, the correction is less than 3%.

**Appendix B: A generalization of the concept of the effective area of the pressure action**

Let's look at water flowing in a Bernoulli tube as described in Figure B1.

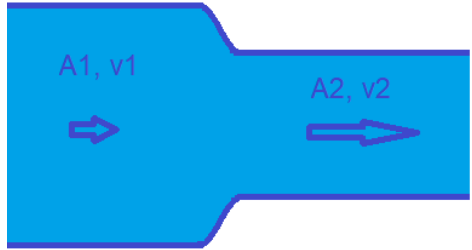

Figure B1: Bernoulli tube

Conservation of flow when transitioning from a wide area to a narrow area yield:

$$(B1) \qquad A_1 v_1 = A_2 v_2$$

Where $A_1$ and $A_2$ represents the cross-sectional area on the left side and on the right side respectively, and $v_1$ and $v_2$ represents the flow velocity on the left side and on the right side respectively. Therefore:

$$(B2) \qquad v_1 = \alpha v_2$$

Where $\alpha$ denotes the ratio of areas: $\alpha = A_2/A_1$.

The average velocity along the acceleration zone is:

$$(B3) \qquad v_{avg} = \frac{v_1 + v_2}{2} = \frac{1+\alpha}{2} v_2$$

By flux conservation we have:

$$A_{avg} v_{avg} = A_2 v_2$$

And thus, the average area of the acceleration zone is:

$$(B4) \qquad A_{avg} = \frac{A_2 v_2}{v_{avg}} = \frac{2}{1+\alpha} A_2 = \frac{2 A_1 A_2}{A_1 + A_2} = \frac{2}{\frac{1}{A_1} + \frac{1}{A_2}}$$

Therefore, we found that $A_{avg}$ is the harmonic mean of $A_1$ and $A_2$.

In the case where $A_2 \ll A_1$, we obtain a factor close to 2: $A_{avg} \cong 2 A_2$, in agreement with the result of chapter 5.

Additionally, according to Bernoulli's equation:

$$(B5) \qquad P_1 + \frac{1}{2}\rho v_1^2 = P_2 + \frac{1}{2}\rho v_2^2$$

Where $P_1$ and $P_2$ represent the pressure at the left side and at the right side respectively.

The recoil force can be calculated in a similar manner to the calculation in section **3.C,** by calculating force accelerating the water according to Bernoulli's law: During the time period $\Delta t$, the water mass $\Delta m = \rho A_2 v_2 \Delta t$ accelerates from velocity $v_1$ to velocity $v_2$. Hence:

$$F_a = \frac{\Delta m \cdot (v_2 - v_1)}{\Delta t} = \rho A_2 v_2 \cdot (v_2 - v_1) = \rho A_2 \cdot (v_2^2 - v_1 v_2) \rightarrow$$

$$(B6) \qquad F_a = \rho A_2 \cdot (1 - \alpha) \cdot v_2^2$$

On the other hand, it can be calculated as the pressure difference multiplied by the average area, $A_{avg}$. Using equation (B4) we get:

$$F_b = A_{avg} \cdot (P_1 - P_2) = A_{avg} \cdot \left(\frac{1}{2}\rho v_2^2 - \frac{1}{2}\rho v_1^2\right) \rightarrow$$

$$(B7) \qquad F_b = \frac{1}{2}\rho \left(\frac{2}{1+\alpha} A_2\right) \cdot (1 - \alpha^2) \cdot v_2^2 = \rho A_2 \cdot (1 - \alpha) \cdot v_2^2$$

The two calculations of recoil force yield the same result. Therefore, we can say that the pressure difference act on an effective area which is the harmonic mean area $A_{avg}$.